\documentclass[%
 aip,
 amsmath,amssymb,
 reprint,%
]{revtex4-1}

\usepackage{graphicx}
\graphicspath{{figures/}}

\usepackage{subcaption}
\usepackage{dcolumn}
\usepackage{bm}

\usepackage[utf8]{inputenc}
\usepackage[T1]{fontenc}
\usepackage{mathptmx}
\usepackage{xcolor}
\usepackage{flafter}
\usepackage{multirow,gensymb}

\newcommand{\blue}[1]{{\color{black}{#1}}}
\newcommand{\blueTwo}[1]{{\color{black}{#1}}}
\begin{document}

\preprint{AIP/123-QED}

\title{Aerosol generation in public restrooms}

\author{Jesse H. Schreck}%
\email{jschreck2015@fau.edu}
\affiliation{Department of Ocean and Mechanical Engineering, Florida Atlantic University, Boca Raton, FL 33431, USA
}%
\author{Masoud Jahandar Lashaki}%
\email{mjahandarlashaki@fau.edu}
\affiliation{Department of Civil, Environmental and Geomatics Engineering, Florida Atlantic University, Boca Raton, FL 33431, USA}
\author{Javad Hashemi}%
\email{jhashemi@fau.edu}
\affiliation{Department of Ocean and Mechanical Engineering, Florida Atlantic University, Boca Raton, FL 33431, USA
}%
\author{Manhar Dhanak}%
\email{dhanak@fau.edu}
\affiliation{Department of Ocean and Mechanical Engineering, Florida Atlantic University, Boca Raton, FL 33431, USA
}%
\author{Siddhartha Verma}
\email{vermas@fau.edu}
\homepage{http://www.computation.fau.edu}
\altaffiliation[Also at ]{Harbor Branch Oceanographic Institute, Florida Atlantic University, Fort Pierce, FL 34946, USA}
\affiliation{Department of Ocean and Mechanical Engineering, Florida Atlantic University, Boca Raton, FL 33431, USA
}%
\date{\today}

\begin{abstract}
Aerosolized droplets play a central role in the transmission of various infectious diseases, including Legionnaire's disease, gastroenteritis-causing norovirus, and most recently COVID-19. Respiratory droplets are known to be the most prominent source of transmission for COVID-19, however, alternative routes may exist given the discovery of small numbers of viable viruses in urine and stool samples. Flushing biomatter can lead to the aerosolization of microorganisms, thus, there is a likelihood that bioaerosols generated in public restrooms may pose a concern for the transmission of COVID-19, especially since these areas are relatively confined, experience heavy foot traffic, and may suffer from inadequate ventilation. To quantify the extent of aerosolization, we measure the size and number of droplets generated by flushing toilets and urinals in a public restroom. The results indicate that the particular designs tested in the study generate a large number of droplets in the size range
$0.3 \mu m$ to $3 \mu m$, which can reach heights of at least $1.52m$. Covering the toilet reduced aerosol levels but did not eliminate them completely, suggesting that aerosolized droplets escaped through small gaps between the cover and the seat. In addition to consistent increases in aerosol levels immediately after flushing, there was a notable rise in ambient aerosol levels due to the accumulation of droplets from multiple flushes conducted during the tests. This highlights the need for incorporating adequate ventilation in the design and operation of public spaces, which can help prevent aerosol accumulation in high occupancy areas and mitigate the risk of airborne disease transmission.
\end{abstract}

\maketitle

\section{Introduction}

The aerosolization of biomatter caused by flushing toilets has long been known to be a potential source of transmission of infectious microorganisms~\cite{Darlow1959,Gerba1975}. Toilet flushing can generate large quantities of microbe-containing aerosols~\cite{Johnson2013} depending on the design and water pressure or flushing energy of the toilet~\cite{Bound1966,Johnson2013Aerosol,Lai2018}. A variety of different pathogens which are found in stagnant water or in waste products (e.g., urine, feces, and vomit) can get dispersed widely via such aerosolization, including the legionella bacterium responsible for causing Legionnaire's disease~\cite{Hamilton2018,Couturier2020}, the Ebola virus~\cite{Lin2017}, the norovirus which causes severe gastroenteritis (food poisoning)~\cite{Caul1994,Marks2000}, and the Middle East Respiratory Syndrome coronavirus (MERS-CoV)~\cite{Zhou2017}. Such airborne dispersion is suspected to have played a key role in the outbreak of viral gastroenteritis aboard a cruise ship, where infection was twice as prevalent among passengers who used shared toilets compared to those who had private bathrooms~\cite{Ho1989}. Similarly, transmission of norovirus via aerosolized droplets was linked to the occurrence of vomiting or diarrhea within an aircraft restroom~\cite{Widdowson2005}, as passengers and crew who got infected were more likely to have visited restrooms than those that were not infected. The participants in the study reported that all of the restroom surfaces appeared to be clean, which indicates that infection is likely to have occurred via bioaerosols suspended within the restroom.

In more controlled studies investigating toilet-generated aerosols, Barker \& Bloomfield~\cite{Barker2000} isolated salmonella bacteria from air samples collected after flushing. Bacteria and viruses could be isolated from settle plates for up to an hour to 90 minutes after flushing~\cite{Barker2005,Best2012}, suggesting that the microorganisms were present in aerosolized droplets and droplet nuclei. An experimental study in a hospital-based setting measured bioaerosol generation when fecal matter was flushed by patients~\cite{Knowlton2018}. A significant increase in bioaerosols was observed right after flushing, and the droplets remained detectable for up to 30 minutes afterwards. Notably, flushing does not remove all of the microorganisms which may be present in the bowl. In various studies where the toilet bowl was seeded with microorganisms, sequential flushes led to a drop in microbe count, however, some residual microbes remained in the bowl even after up to 24 flushes~\cite{Gerba1975,Barker2000,Barker2005,Johnson2017,Aithinne2019}. In some cases, residual microbial contamination was shown to persist in biofilm formed within the toilet bowl for several days to weeks~\cite{Barker2000}. 

In an effort to reduce aerosol dispersal, certain studies conducted measurements with the toilet seat lid closed~\cite{Barker2005,Best2012}. Closing the lid led to a decrease, but not a complete absence of bacteria recovered from air samples. This suggests that smaller aerosolized droplets were able to escape through the gap between the seat and the lid. In addition to the experiment-based studies mentioned here, numerical simulations have been used recently to investigate the ejection of aerosolized particles from toilets and urinals, specifically in the context of COVID-19 transmission~\cite{Li2020,Wang2020}.

The issue of aerosolization is particularly acute for viruses compared to bacteria, given their different response to levels of relative-humidity (RH). High RH levels result in slower evaporation of aerosolized droplets, whereas lower levels accelerate the phenomenon, leading to the formation of extremely small droplet nuclei which can remain airborne for long periods of time\blue{ and can deposit deep into the lungs~\cite{Mallik2020,Wang2020Motion}}. Various studies have indicated that the viability of bacteria decreases at low RH levels~\cite{Won1966,Lin2020}, which makes them less likely to retain their infectivity in droplet nuclei form. On the other hand, viruses exhibit lowest viability at intermediate RH levels, and retain their viability at either low or high RH values~\cite{Songer1967,Benbough1971,Schaffer1976,Donaldson1976,Lin2020}, making them more likely to remain intact in droplet nuclei which can stay suspended from hours to days. Viruses are also more likely to aerosolize easily, as indicated by Lee at al.~\cite{Lee2016} who used wastewater sludge (both synthetic and real) to demonstrate that when viruses were seeded into the sludge, $94\%$ stayed mobile in the liquid phase while only a small fraction adhered to the solid biomatter or to the surfaces of the toilet. This suggests that the presence of solid biomatter, which is more difficult to aerosolize, might not reduce the potential for virus transmission since they are more likely to get aerosolized with the liquid phase.

Apart from gastrointestinal diseases, viruses associated with respiratory illnesses have also been detected in patients' stool and urine samples. For instance, the SARS-CoV (Severe Acute Respiratory Syndrome Coronavirus) responsible for the SARS outbreak of 2003 was found in patients' urine and stool specimens for longer than 4 weeks~\cite{Xu2005}. Similarly, recent studies have confirmed the presence of SARS-CoV-2 (the virus associated with COVID-19) viral RNA in patients' stool samples~\cite{Xiao2020,Wu2020Lancet,Chen2020,Xiao2020,Zhang2020Emerging,Foladori2020,Gupta2020}, even if they did not experience gastrointestinal symptoms and regardless of the severity of their respiratory symptoms~\cite{Wu2020Lancet,Chen2020,Zhang2020,Ling2020}. Surprisingly, viral RNA could be detected in feces for several days to weeks after it was no longer detectable in respiratory samples from nasal and oral swabs~\cite{Xiao2020,Wu2020Lancet,Chen2020,Gupta2020}. Moreover, Wu et al.~\cite{Wu2020} recovered large quantities of viral RNA from urban wastewater treatment facilities. The levels detected were several orders of magnitude higher than would be expected for the number of clinically confirmed cases in the region, which suggests that there was a high prevalence of asymptomatic and undetected cases.

Although enveloped viruses like SARS-CoV-2 are susceptible to the acids and bile salts found in digestive juices, it has been shown that they can survive when engulfed within mucus produced by the digestive system. Hirose et al.~\cite{Hirose2017} demonstrated that influenza viruses could be protected from degradation by simulated digestive juices using both artificial and natural mucus. This might help explain why recent studies have been able to isolate viable SARS-CoV-2 virus particles (i.e., those able to infect new cells) that remained intact when passing through the digestive and urinary systems, albeit in smaller quantities compared to respiratory fluids~\cite{Jones2020}.  Wang et al.~\cite{Wang2020JAMA} detected live virus in feces from patients who did not have diarrhea, and Xiao et al.~\cite{Xiao2020Viable} demonstrated the infectivity of intact virions isolated from a patient's stool samples. In urine specimens, SARS-CoV-2 RNA is found less frequently than in fecal and respiratory samples~\cite{Peng2020,Ling2020,Xiao2020}. However, Sun et al.~\cite{Sun2020} managed to isolate the virus from a severely infected patient's urine, and showed that these virions were capable of infecting new susceptible cells. As with fecal samples, viral RNA has been found in urine even after the virus is no longer detectable in respiratory swabs~\cite{Ling2020}.

These findings suggest that the aerosolization of biomatter could play a potential role in the transmission of SARS-CoV-2, which is known to remain viable in aerosol form~\cite{vanDoremalen2020,Fears2020}. Environmental samples taken by Ding et al.~\cite{Ding2020} in a hospital designated specifically for COVID-19 patients indicated high prevalence of the virus within bathrooms used by the patients, both on surfaces and in air samples. The authors hypothesized that aerosolized fecal matter may have dispersed the virus within the bathroom, since viral samples were not detected on surfaces in the patients' rooms. 

Given the potential role of aerosolized biomatter in spreading a wide variety of gastrointestinal and respiratory illnesses, we investigate droplet generation from toilets and urinals in a public restroom operating under normal ventilation condition. We examine the size, number, and various heights to which the droplets rise when generated by the flushing water. The main aim is to better understand the risk of infection transmission that the droplets pose in public restrooms, since these relatively confined locations often experience heavy foot traffic. The experimental methodology is described in Section~\ref{sec:methods} followed by results and discussion in Section~\ref{sec:results} and conclusion in Section~\ref{sec:conclusion}.

\section{Methods}
\label{sec:methods}
The flush-generated aerosol measurements were recorded in a medium sized restroom on the university campus, consisting of 3 bathroom cubicles, 6 urinals, and 3 sinks. The restroom was deep cleaned and closed twenty four hours prior to conducting the experiments, with the ventilation system operating normally to remove any aerosols generated during cleaning. \blue{The temperature and relative humidity within the restroom were measured to be $21\degree C$ and $52\%$, respectively.} For the measurements reported here, one particular toilet and one urinal were selected, both equipped with flushometer type flushing systems. The urinal used 3.8 liters of water per flush whereas the toilet used 4.8 liters per flush. 

The size and concentration of aerosols generated by flushing were measured using a handheld particle counter (9306-V2 - TSI Incorporated). \blue{The sensor's size resolution is less than $15\%$, which is indicative of the uncertainty in the measured particle diameter. More specifically, the resolution is specified as the ratio of standard deviation to the mean size of the particles being sampled. The counting efficiency of the sensor is $50\%$ at $0.3\mu m$ and $100\%$ for particles larger than $0.45\mu m$. These values denote the ratio of particle numbers measured by the counter to those measured using a reference instrument. Handheld counters with comparable specifications have been used for estimating the likelihood of aerosol transmission in typical public spaces~\cite{Somsen2020}.} 

\begin{figure*}[ht!]
\centering
\begin{subfigure}{0.38\linewidth}
\centering
\includegraphics[width=\linewidth]{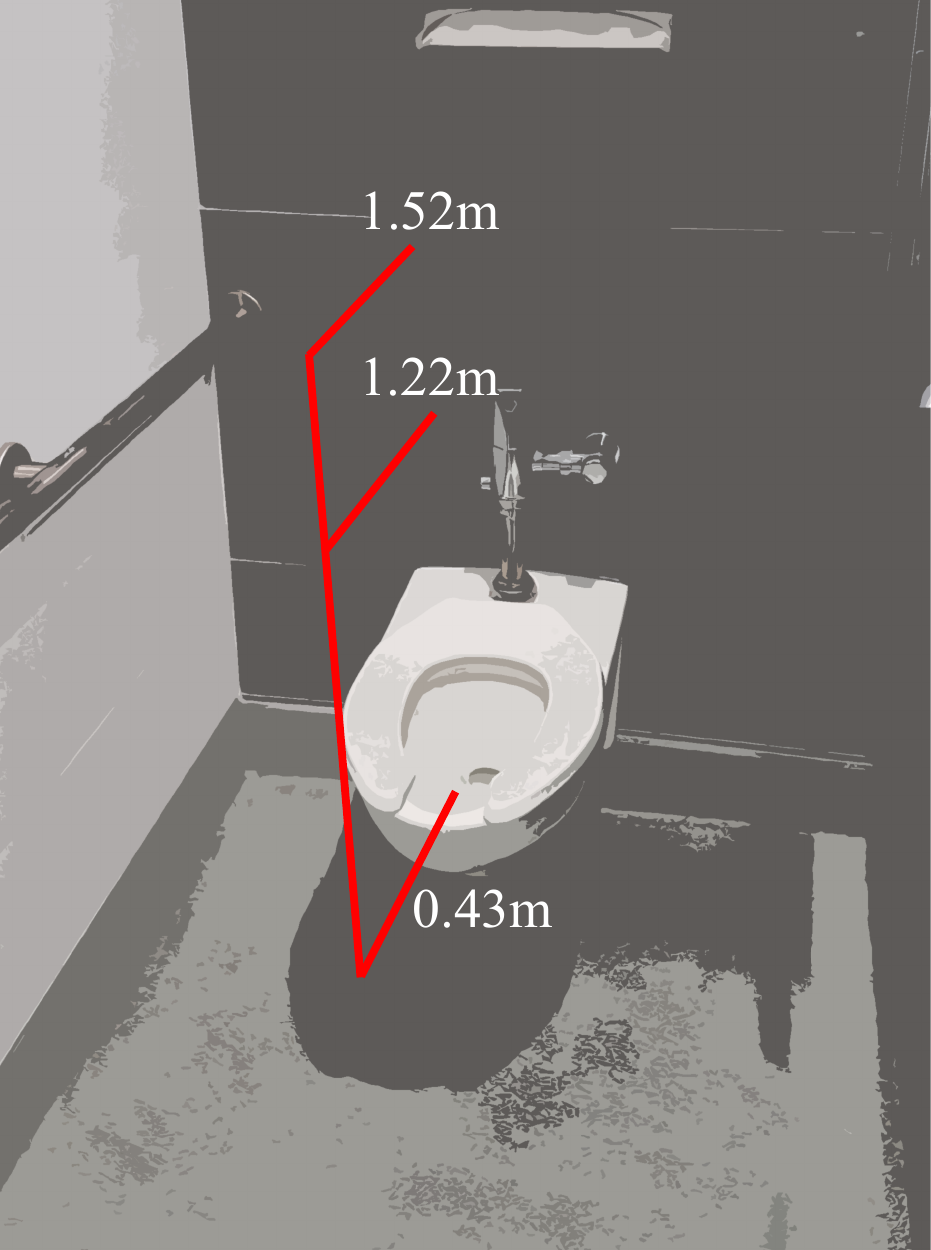}
\caption{\label{fig:toiletSchemat}}
\end{subfigure}
\begin{subfigure}{0.379\linewidth}
\centering
\includegraphics[width=\linewidth]{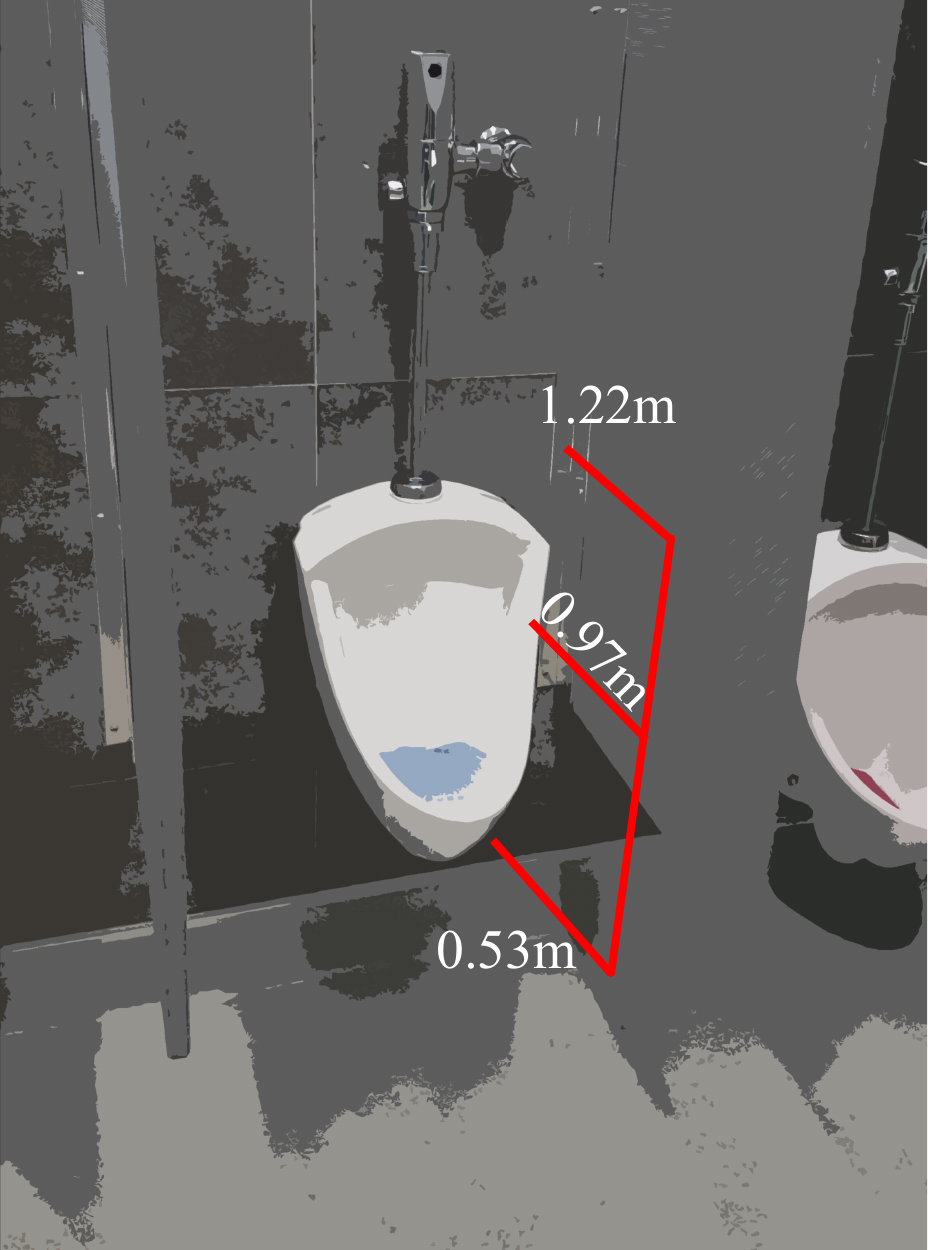}
\caption{\label{fig:urinalSchemat}}
\end{subfigure}
\caption{\label{fig:Schematics} Measurement locations where the aerosol sensor was placed for (\subref{fig:toiletSchemat}) the toilet and (\subref{fig:urinalSchemat}) the urinal. Measurements for the toilet were taken at heights of $0.43m$ from the ground ($1ft \ 5in$), $1.22m$ ($4ft$), and  $1.52m$ ($5ft$), whereas those for the urinal were taken at $0.53m$ ($1ft \ 9in$), $0.97m$ ($3ft \ 2in$), and $1.22m$ ($4ft$).}
\end{figure*}

The \blue{particle} counter was positioned at various heights close to the toilet and the urinal as shown in Figure~\ref{fig:Schematics}.
Measurements for the toilet were taken at 3 different heights, at approximately $0.43m$ from the ground ($1ft \ 5in$), $1.22m$ ($4ft$), and $1.52m$ ($5ft$), with the toilet seat raised up. The lowest level corresponds to the distance between the ground and the toilet seat, and represents the scenario where the particle counter was placed level with the seat. Measurements for the urinal were taken at \blue{3 different heights, at approximately $0.53m$ from the ground ($1ft \ 9in$), $0.97m$ ($3ft \ 2in$), and $1.22m$ ($4ft$)}. \blue{The particle counter's intake probe was oriented parallel to the floor and perpendicular to the back wall, with the inlet pointing in the direction of the flushing water. The probe was centered laterally for both toilet and urinal measurements. The placement and orientation were selected to be representative of a person breathing in when flushing the toilet/urinal after use, since different choices were observed have a notable impact on the measured droplet count. The probe inlet was positioned $5cm$ inside the rim of the toilet, and it was placed $5cm$ outside the edge of the urinal, as depicted in Figure~\ref{fig:Schematics}.} In addition to measurements taken during normal operation of the toilet, aerosol measurements were recorded after a large flat plate was placed over the toilet opening, to assess the impact of flushing with the lid closed. The use of a separate cover was necessary since public restrooms in the United States often do not come equipped with toilet seat lids.

The particle counter drew air samples at a volume flow rate of 2.83 liters per minute \blue{(0.1 Cubic Feet per Minute - CFM)}, and measured aerosol concentrations in six different size ranges, namely, (0.3 to 0.5)$\mu m$, (0.5 to 1.0)$\mu m$, (1.0 to 3.0)$\mu m$, (3.0 to 5.0)$\mu m$, (5.0 to 10.0)$\mu m$, and (10.0 to 25.0)$\mu m$. For the tests reported here, air samples were recorded at a sampling frequency of $1Hz$ for a total of 300 seconds at each of the levels depicted in Figure~\ref{fig:Schematics}. \blue{We note that although it is feasible to compute droplet concentration at a given measurement location, it is difficult to determine overall characteristic droplet production rates for the toilet or urinal, since the measured values depend on both the location and orientation of the probe.} During the 300-second sampling, the toilet and urinal were flushed manually 5 different times at the 30, 90, 150, 210, and 270 second mark, with the flushing handle held down for five consecutive seconds. The data obtained from the three different scenarios, i.e., toilet flushing, covered toilet flushing, and urinal flushing, were analyzed to determine the increase in aerosol concentration. The behavior of droplets of different sizes, the heights that they rose to, and the impact of covering the toilet are discussed in detail in Section~\ref{sec:results}.

\section{Results and Discussion}
\label{sec:results}
\begin{figure*}[ht]
\centering
\begin{subfigure}{0.49\linewidth}
\centering
\includegraphics[width=\linewidth]{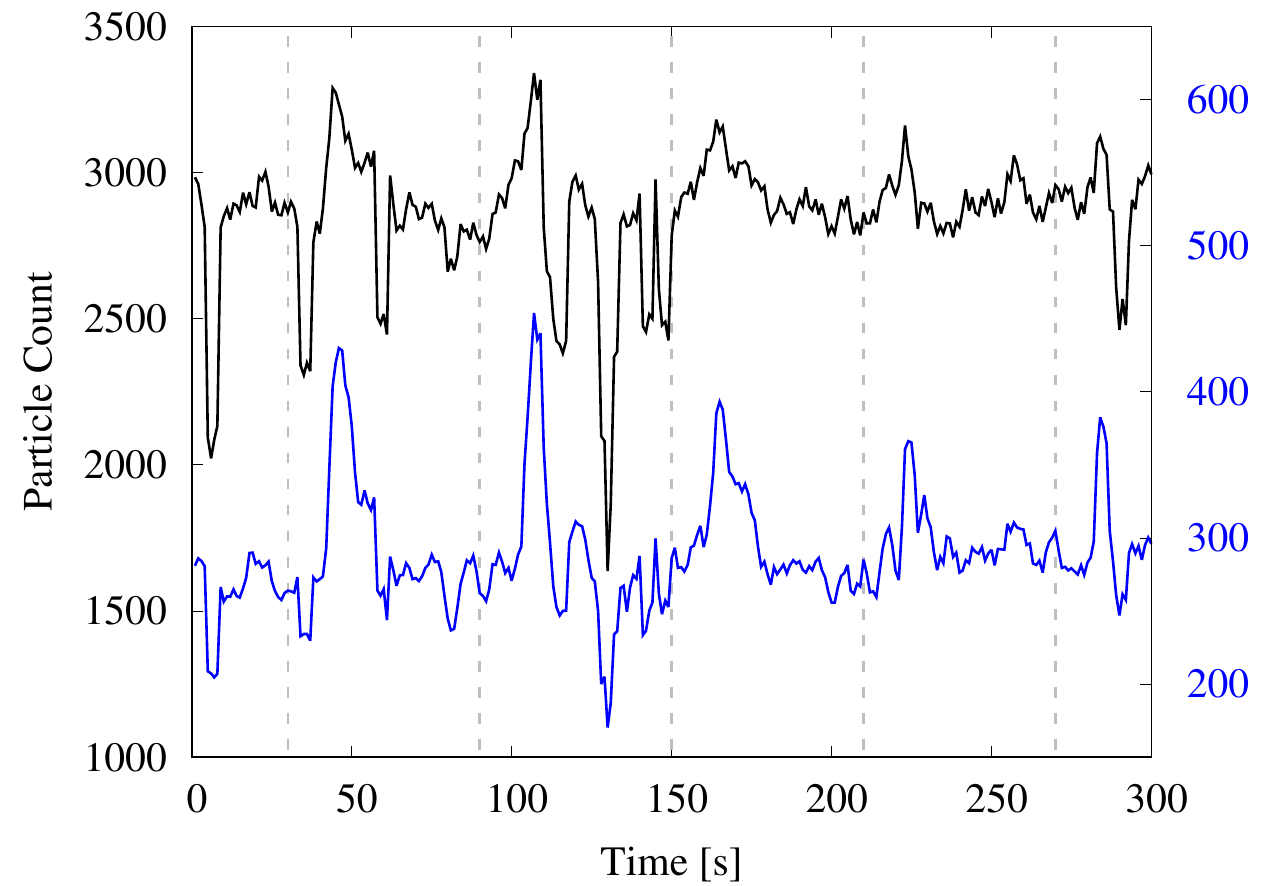}
\caption{\label{fig:T_pt3topt5andpt5to1}}
\end{subfigure}
\begin{subfigure}{0.49\linewidth}
\centering
\includegraphics[width=\linewidth]{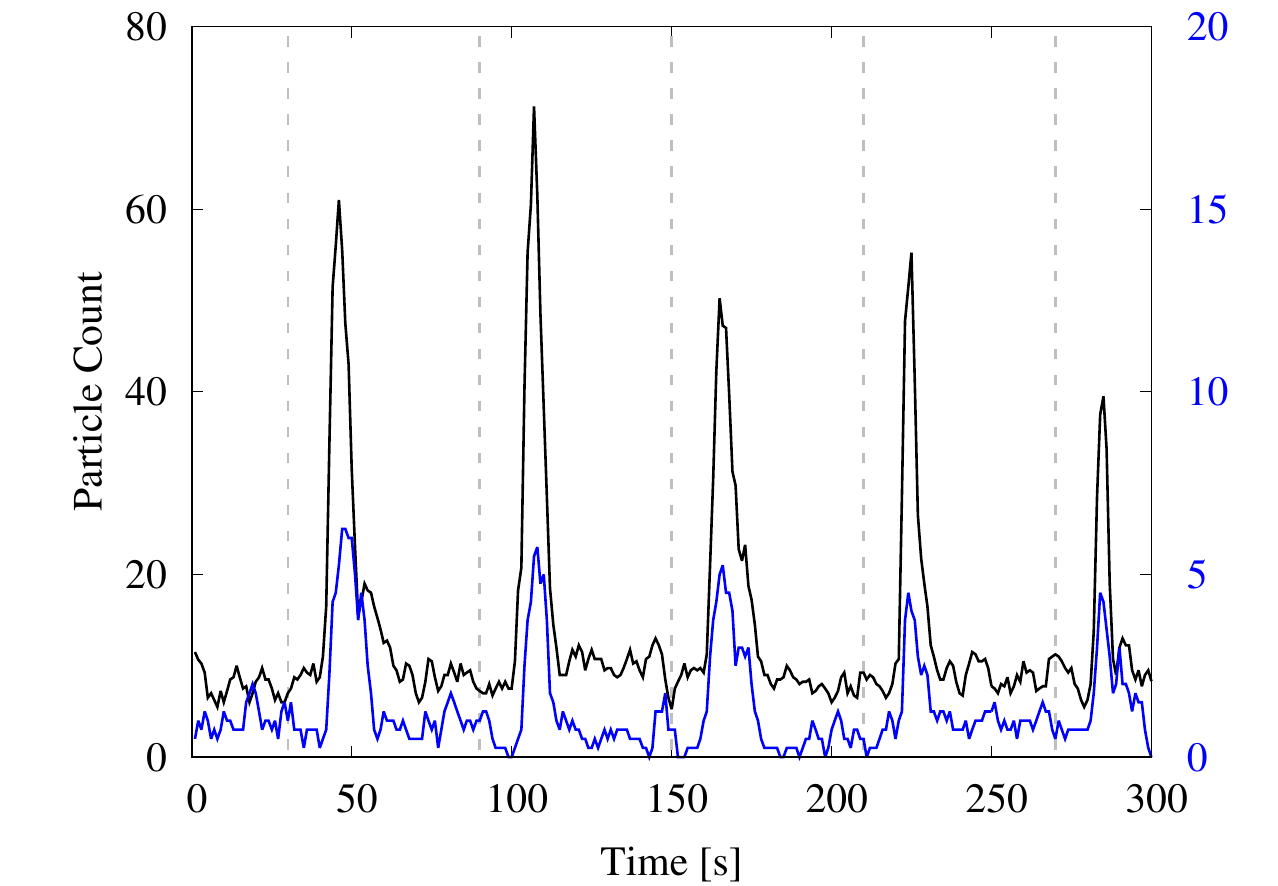}
\caption{\label{fig:T_1to3and3to5}}
\end{subfigure}
\begin{subfigure}{0.49\linewidth}
\centering
\includegraphics[width=\linewidth]{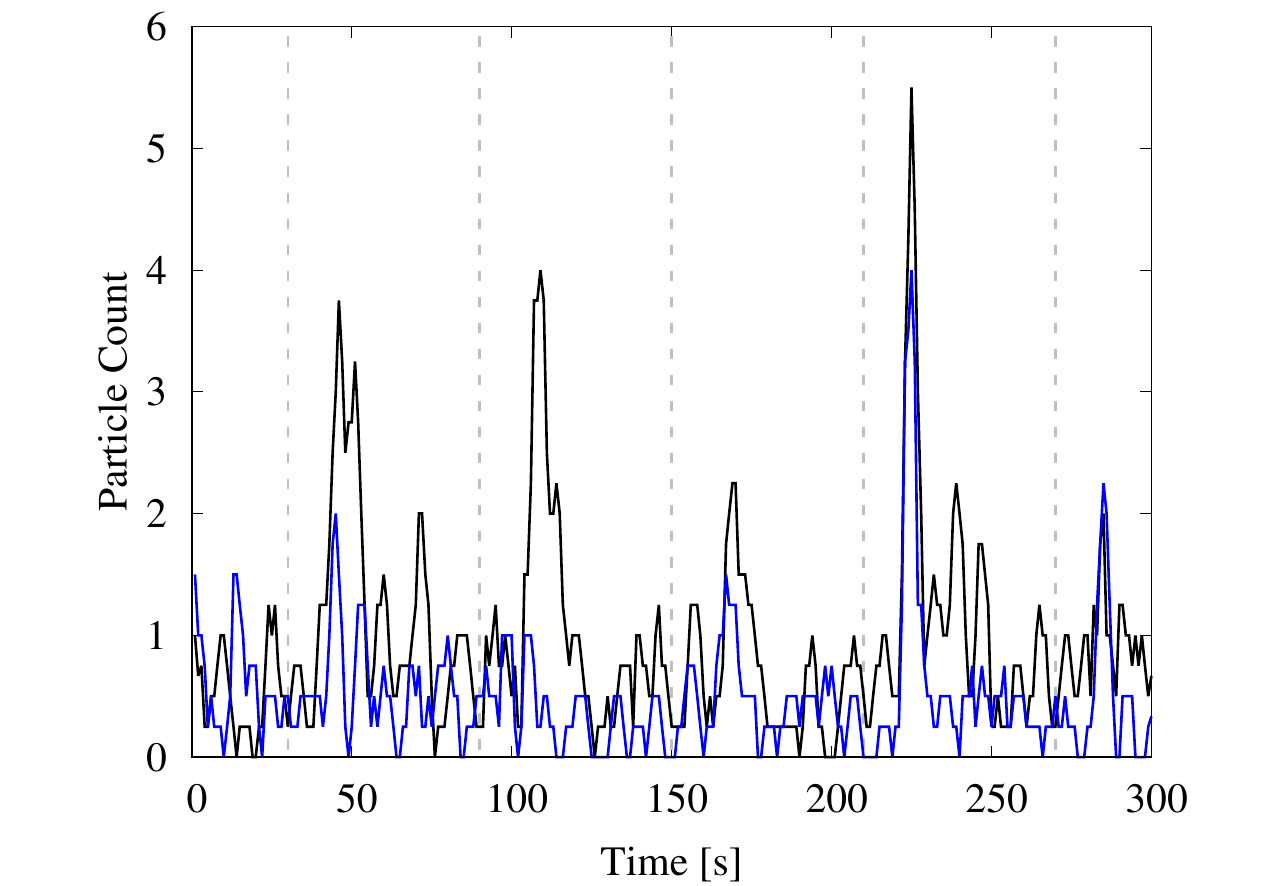}
\caption{\label{fig:T_5to10and10to25}}
\end{subfigure}
\caption{\label{fig:ToiletTime}Particle-count from the toilet-flushing test, measured at a height of $0.43m$ ($1ft 5in$). The time series plots are shown for particles in various size ranges: (\subref{fig:T_pt3topt5andpt5to1}) (0.3 to 0.5)$\mu m$ - black, and (0.5 to 1)$\mu m$ - blue; (\subref{fig:T_1to3and3to5}) (1 to 3)$\mu m$ - black, and (3 to 5)$\mu m$ - blue; (\subref{fig:T_5to10and10to25}) (5 to 10)$\mu m$ - black, and (10 to 25)$\mu m$ - blue.  The black curves in (\subref{fig:T_pt3topt5andpt5to1}) and (\subref{fig:T_1to3and3to5}) correspond to the left vertical axes, whereas the blue curves correspond to the right vertical axes. The dashed gray lines indicate the instances when the flushing handle was depressed and held down for 5 seconds.}
\end{figure*}

The measurements from the particle counter were analyzed to determine the extent of aerosolization, and the various heights to which the droplets rise after flushing. Figure~\ref{fig:ToiletTime} shows the time-variation of the total number of particles recorded by the sensor from measurements for the uncovered toilet.
The data plotted has been smoothed using a moving average window of size 4 to reduce noise levels. Figure~\ref{fig:T_pt3topt5andpt5to1} depicts the time series for particles of size (0.3 to 0.5)$\mu m$ and (0.5 to 1)$\mu m$, whereas the size groups (1 to 3)$\mu m$ and (3 to 5)$\mu m$ are shown in Figure~\ref{fig:T_1to3and3to5}, and size groups (5 to 10)$\mu m$ and (10 to 25)$\mu m$ are shown in Figure~\ref{fig:T_5to10and10to25}. We observe a noticeable increase in particle count for all of the size ranges a few seconds after flushing. This indicates that flushing the toilet generates droplets \blue{in significant numbers}, which can be detected at seat-level for up to 30 seconds after initiating the flush.

In Figure~\ref{fig:T_pt3topt5andpt5to1} we observe a large variation in the measured levels of the smallest particles, i.e., those smaller than $1 \mu m$. These particles are highly susceptible to flow disturbances in the ambient environment due to their low mass, which may account for the high variability. The time series for particles larger than 1$\mu m$ (Figures~\ref{fig:T_1to3and3to5} and~\ref{fig:T_5to10and10to25}) exhibit distinctive surges in particle count after each flushing event. Importantly, the total number of droplets generated in the smaller size ranges is considerably larger than that generated in the larger ranges, even though the surges appear to be less prominent for the smaller droplets. We note that for the smallest aerosols (i.e., those smaller than 1$\mu m$), ambient levels in the restroom were relatively high prior to starting the experiment ($\sim O(3000)$). Thus, in these size ranges the flush-generated droplets comprise a small fraction of the total particle count. On the other hand, ambient levels for particle sizes larger than 1$\mu m$ were negligible in the restroom ($\sim O(1)$ to $O(10)$), resulting in the distinctive surges observed after flushing. 

\begin{figure*}[ht!]
\centering
\begin{subfigure}{0.49\linewidth}
\centering
\includegraphics[width=\linewidth]{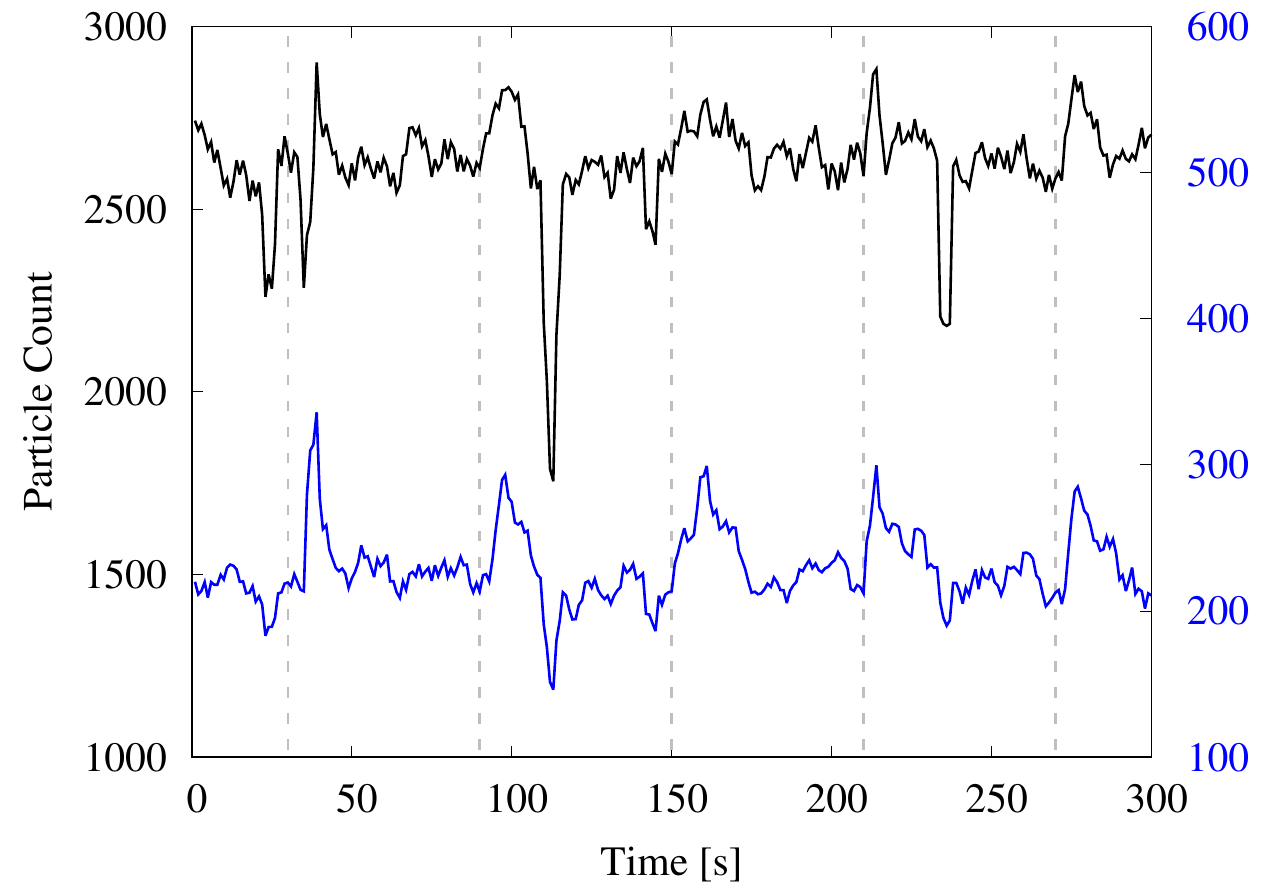}
\caption{\label{fig:TC_pt3topt5andpt5to1}}
\end{subfigure}
\begin{subfigure}{0.49\linewidth}
\centering
\includegraphics[width=\linewidth]{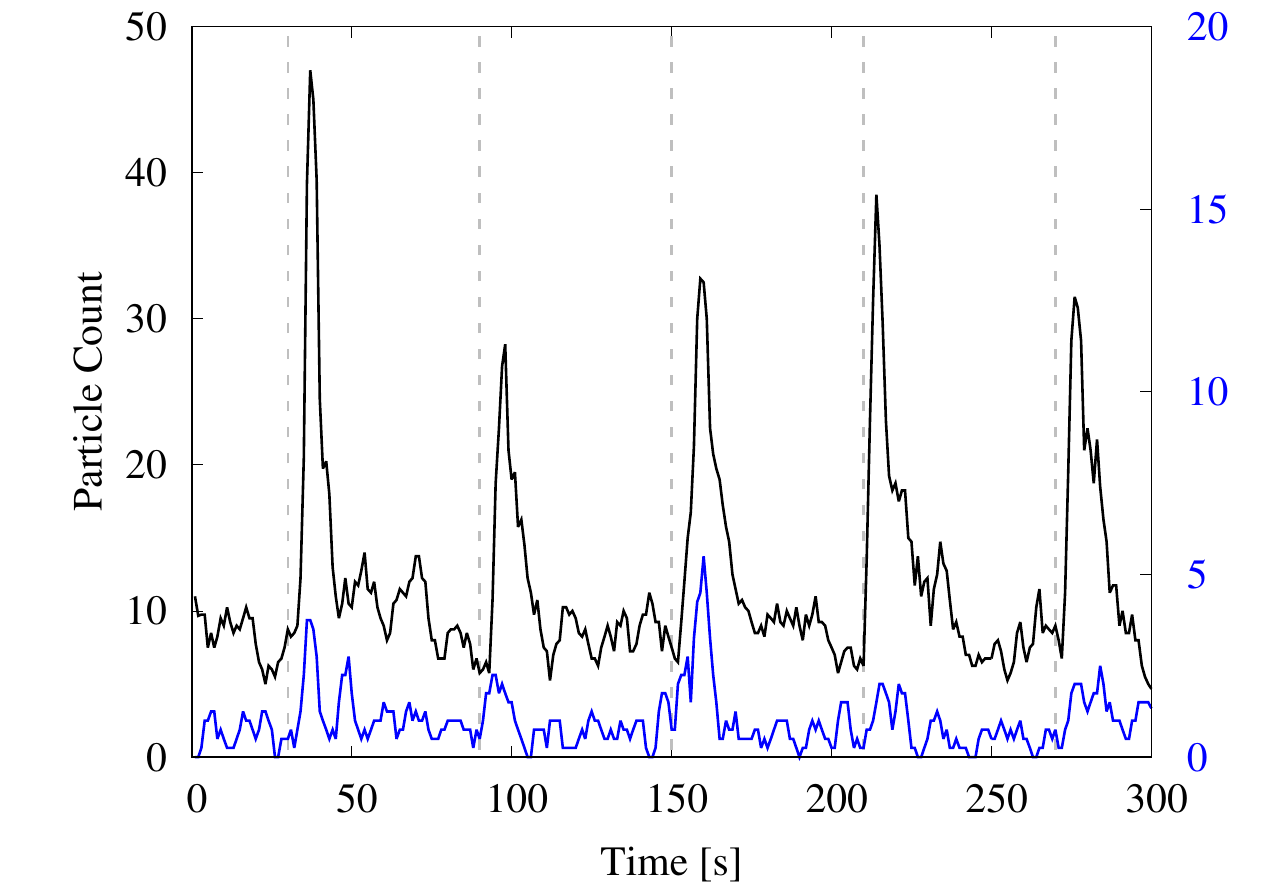}
\caption{\label{fig:TC_1to3and3to5}}
\end{subfigure}
\begin{subfigure}{0.49\linewidth}
\centering
\includegraphics[width=\linewidth]{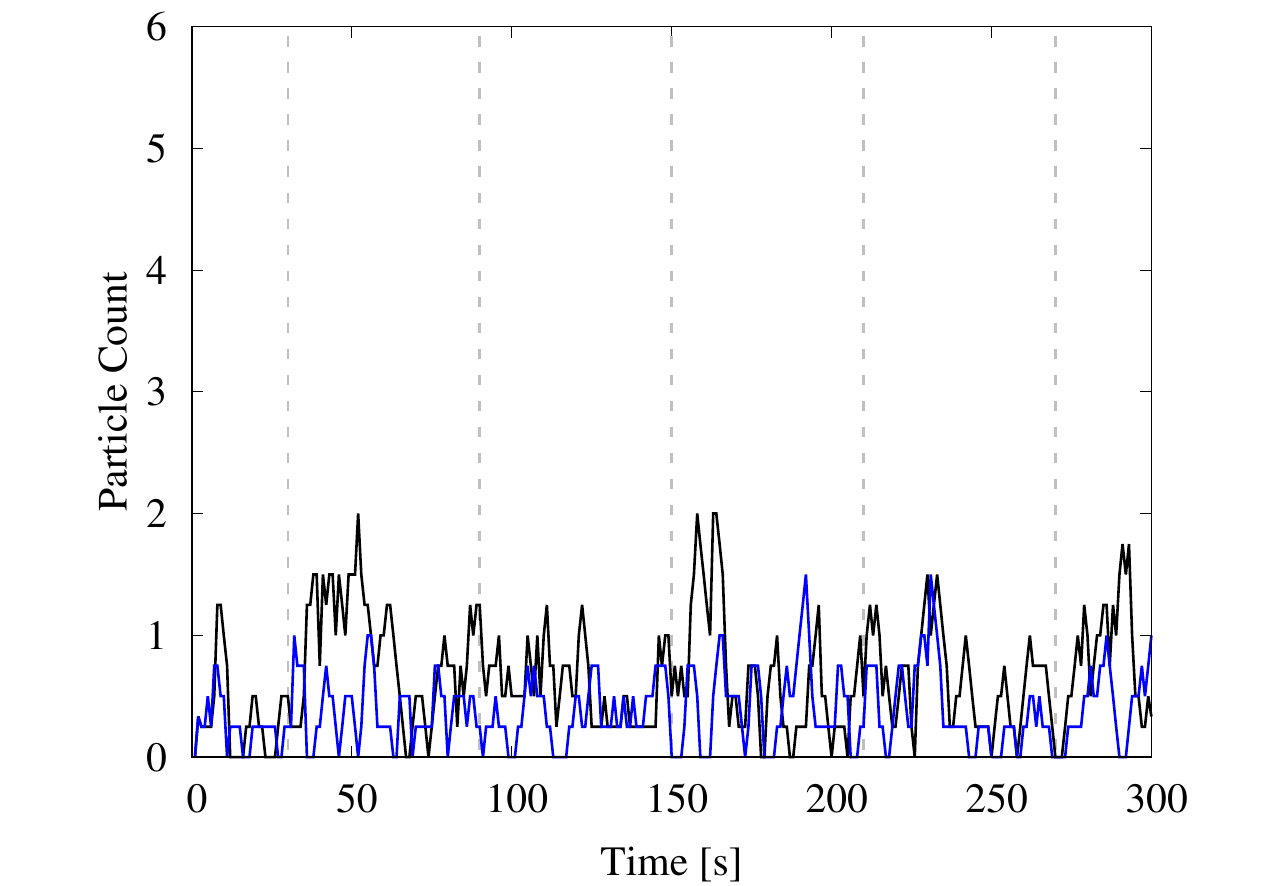}
\caption{\label{fig:TC_5to10and10to25}}
\end{subfigure}
\caption{\label{fig:CovToiletTime}Particle-count from the flushing test when the toilet was covered using a large flat plate. Measurements taken at a height of $0.43m$ ($1ft 5in$). The time series plots are shown for particles in various size ranges: (\subref{fig:TC_pt3topt5andpt5to1}) (0.3 to 0.5)$\mu m$ - black, and (0.5 to 1)$\mu m$ - blue; (\subref{fig:TC_1to3and3to5}) (1 to 3)$\mu m$ - black, and (3 to 5)$\mu m$ - blue; (\subref{fig:TC_5to10and10to25}) (5 to 10)$\mu m$ - black, and (10 to 25)$\mu m$ - blue.  The black curves in (\subref{fig:TC_pt3topt5andpt5to1}) and (\subref{fig:TC_1to3and3to5}) correspond to the left vertical axes, whereas the blue curves correspond to the right vertical axes. The dashed gray lines indicate the instances when the flushing handle was depressed and held down for 5 seconds.}
\end{figure*}

\begin{figure*}[ht]
\centering
\begin{subfigure}{0.49\linewidth}
\centering
\includegraphics[width=\linewidth]{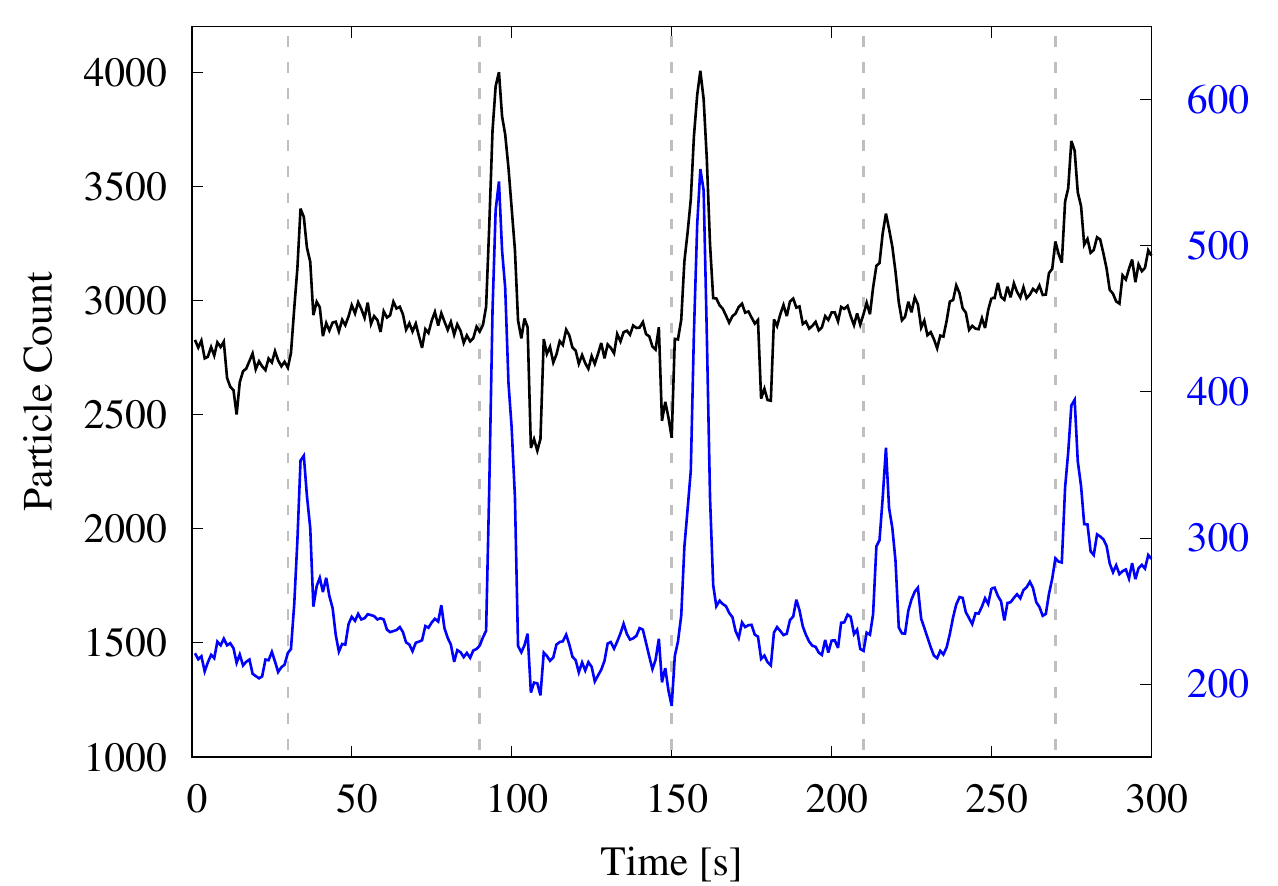}
\caption{\label{fig:U_pt3topt5andpt5to1}}
\end{subfigure}
\begin{subfigure}{0.48\linewidth}
\centering
\includegraphics[width=\linewidth]{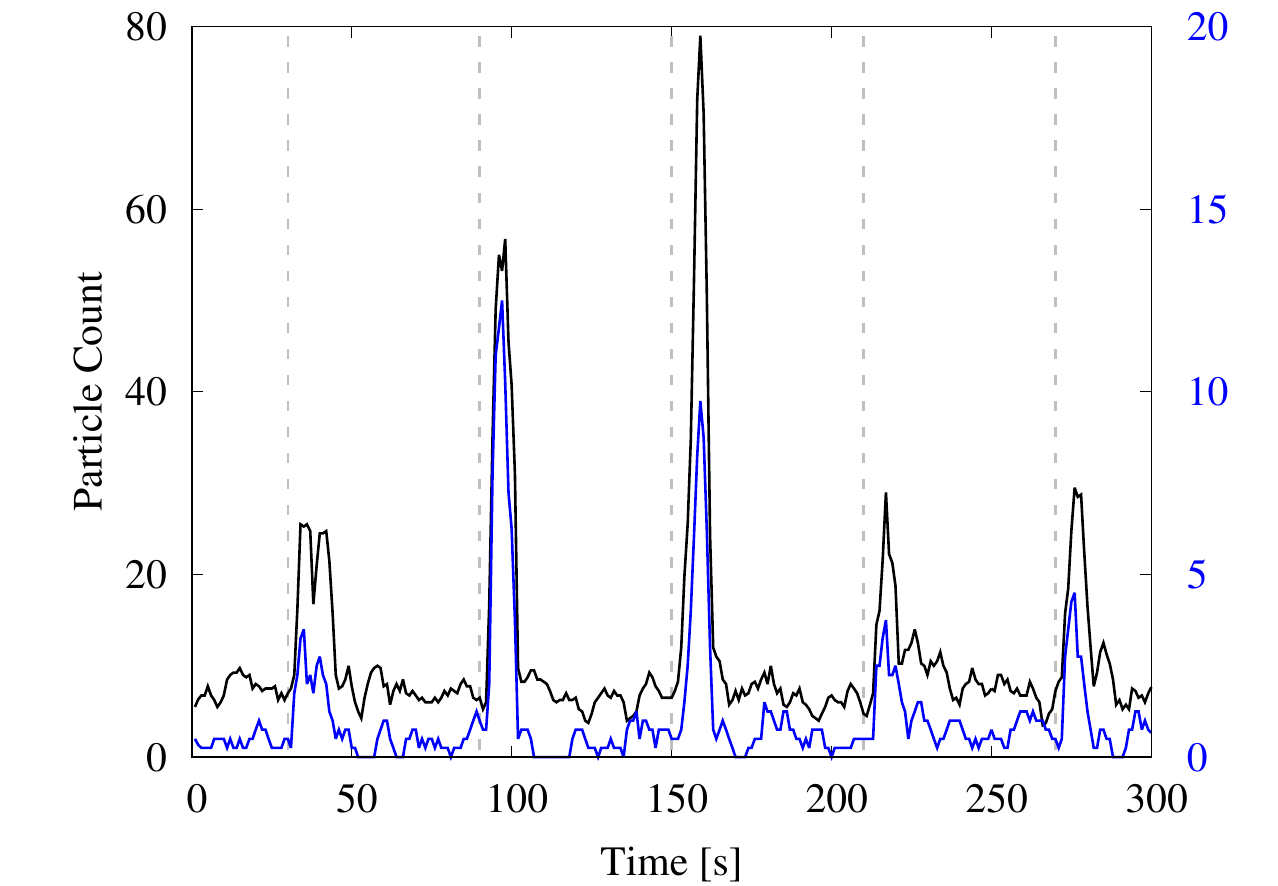}
\caption{\label{fig:U_1to3and3to5}}
\end{subfigure}
\begin{subfigure}{0.48\linewidth}
\centering
\includegraphics[width=\linewidth]{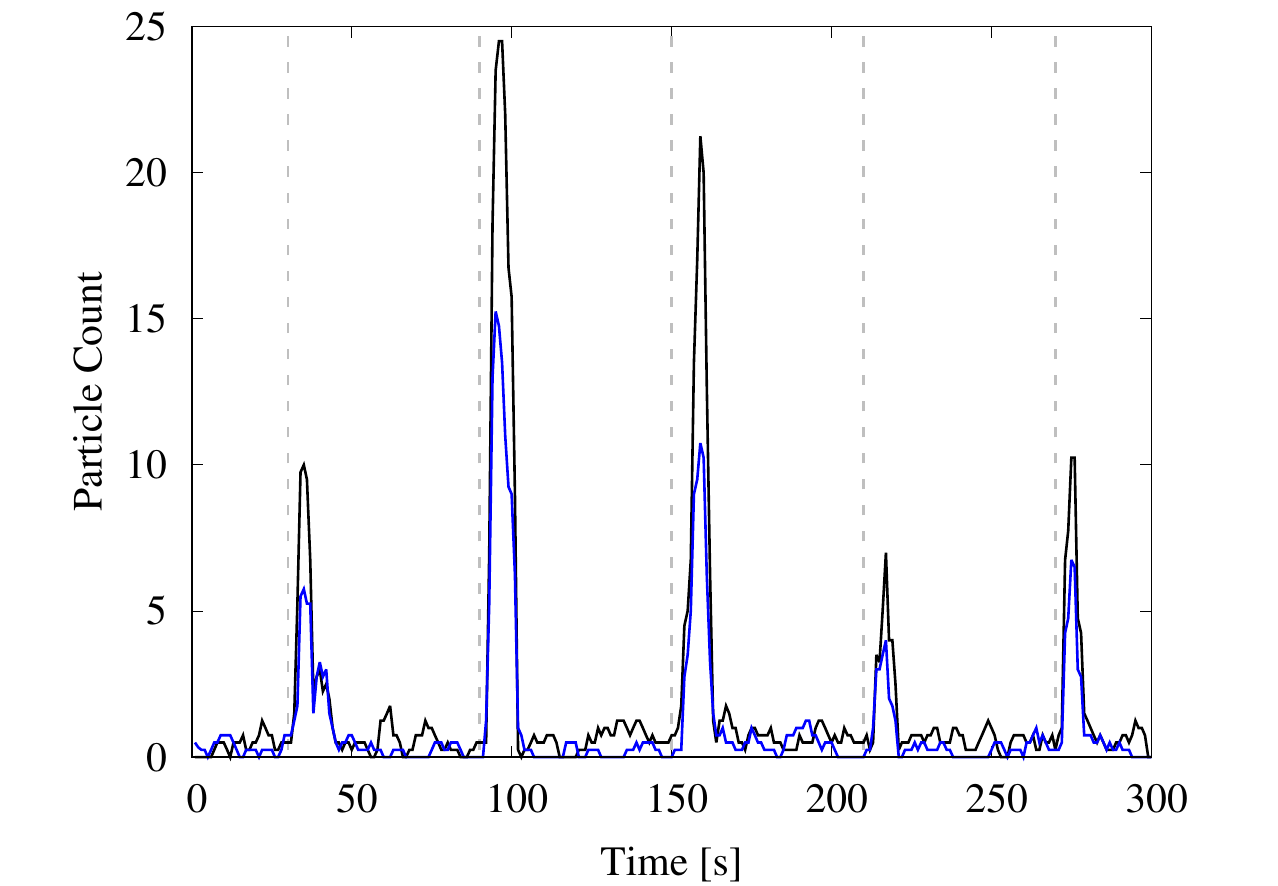}
\caption{\label{fig:U_5to10and10to25}}
\end{subfigure}
\caption{\label{fig:UrinalTime}Particle-count from the urinal-flushing test, measured at a height of $0.53m$ ($1ft 9in$). The time series plots are shown for particles in various size ranges: (\subref{fig:U_pt3topt5andpt5to1}) (0.3 to 0.5)$\mu m$ - black, and (0.5 to 1)$\mu m$ - blue; (\subref{fig:U_1to3and3to5}) (1 to 3)$\mu m$ - black, and (3 to 5)$\mu m$ - blue; (\subref{fig:U_5to10and10to25}) (5 to 10)$\mu m$ - black, and (10 to 25)$\mu m$ - blue. The black curves in (\subref{fig:U_pt3topt5andpt5to1}) and (\subref{fig:U_1to3and3to5}) correspond to the left vertical axes, whereas the blue curves correspond to the right vertical axes. The dashed gray lines indicate the instances when the flush was activated using the proximity sensor.}
\end{figure*}

Similar plots depicting the time-variation of droplet counts for the covered toilet test and the urinal-flushing test are shown in Figure~\ref{fig:CovToiletTime} and Figure~\ref{fig:UrinalTime}, respectively.
For the covered toilet, the plots display a large variation in the number of the smallest droplets in Figure~\ref{fig:TC_pt3topt5andpt5to1}, and comparatively small surges relative to ambient levels due to the background count being high. Importantly, the observed peak values of the surges are lower for the covered toilet compared to the uncovered tests. This is evident in Figure~\ref{fig:TC_1to3and3to5}, where the peak values are approximately 35 droplets on average for the (1 to 3)$\mu m$ range, and 3 droplets for the (3 to 5)$\mu m$ range. The same numbers for the uncovered toilet are approximately 50 droplets and 5 droplets, respectively, in Figure~\ref{fig:T_1to3and3to5}. Notably, there is a significant reduction in the number of droplets larger than 5$\mu m$ for the covered toilet (Figure~\ref{fig:TC_5to10and10to25}) compared to the uncovered toilet (Figure~\ref{fig:T_5to10and10to25}). This indicates that the covering helps to reduce the dispersion of flush-generated droplets, especially those larger than 5$\mu m$, but it does not completely contain the escape of droplets smaller than 5$\mu m$.

The data from the urinal-flushing tests in Figure~\ref{fig:UrinalTime} indicate a large number of droplets generated in all size ranges observed; the post-flush surges are much more pronounced than those for the toilet-flushing tests, even for droplets smaller than 1$\mu m$ (Figure~\ref{fig:U_pt3topt5andpt5to1}). This may be related to the closer proximity of the sensor to the water drain in the urinal, compared to the toilet-flushing tests for which the sensor was placed at the outer edge of the toilet bowl. \blue{We observe that there is no consistent increasing or decreasing trend in either the peaks or the baseline levels with subsequent flushes in the time series plots. The same holds true for data from the toilet-flushing tests in Figures~\ref{fig:ToiletTime} and \ref{fig:CovToiletTime}. Thus, any short term changes in temperature and RH at the measurement location due to flushing do not have a noticeable impact on the droplet count.} \blue{Furthermore, while the smallest droplets will remain suspended for longer than $300s$, the time series plots indicate that  droplet counts at the sensor location return to ambient levels within approximately half a minute. Nonetheless, as these droplets move past the particle counter they become part of the ambient environment, leading to a measurable increase in background levels as demonstrated later in this section.}

To compare the increase in droplet concentration for the three different scenarios at various measurement heights, the time series data were examined manually to identify the time delay between flush initiation and the observed rise in particle count, as well as the total time span for which the particle counts remained elevated. The corresponding values are provided in Table~\ref{tab:startAndSpan}.
\begin{table}[htp]
	\caption{\label{tab:startAndSpan} Average time delay between flush initiation and the observed rise in particle count. The last column indicates the average time taken for the particle count to return to ambient levels.}
\blue{
\begin{ruledtabular}
\begin{tabular}{llcc}
	&Height 	&Time Delay [s]  &Time Span [s]  \\ \hline
    \multirow{3}{*}{Toilet} 
    &$0.43m$ ($1ft 5in$)  	& 10		& 20 \\
	&$1.22m$ ($4ft$)   		& 10		& 20 \\
	&$1.52m$ ($5ft$)      	& 10		& 20 \\
    \hline    
    \multirow{2}{*}{Covered Toilet} 
    &$0.43m$ ($1ft 5in$) 	& 0		& 20 	\\
	&$1.22m$ ($4ft$) 		& 5		& 20 	\\
    \hline
    \multirow{2}{*}{Urinal}
    &$0.53m$ ($1ft 9in$) 	& 0		& 15 \\
	&$0.97m$ ($3ft 2in$) 	& 5		& 15 \\
	&$1.22m$ ($4ft$)    	& 6		& 20
\end{tabular}
\end{ruledtabular}
}
\end{table}
We note that the time delay between flush initiation and the measured surge for the uncovered toilet at seat-level was 10 seconds, whereas that for the covered toilet was 0 seconds. \blue{Furthermore, the delay was smaller for the covered toilet at a height of $1.22m$ ($5s$ versus $10s$), suggesting that the aerosols were forced through gaps in between the seat and the plate for the covered toilet}. In both cases, the droplet counts remained elevated for a further 20 seconds after first detection of the surge. \blue{For the covered toilet and the urinal, we observe a consistent increase in time delay with increasing height, which also corresponds to increasing distance from the flushing water, but the observed delay remained nearly constant for the uncovered toilet at $10s$}. \blue{We remark that the time delay and detection duration are expected to be influenced strongly} by the placement of the sensor, the \blue{fixture} geometry, the flushing mechanism, as well as the water volume \blue{and pressure}.

\begin{figure}[h!]
\centering
\includegraphics[width=\linewidth]{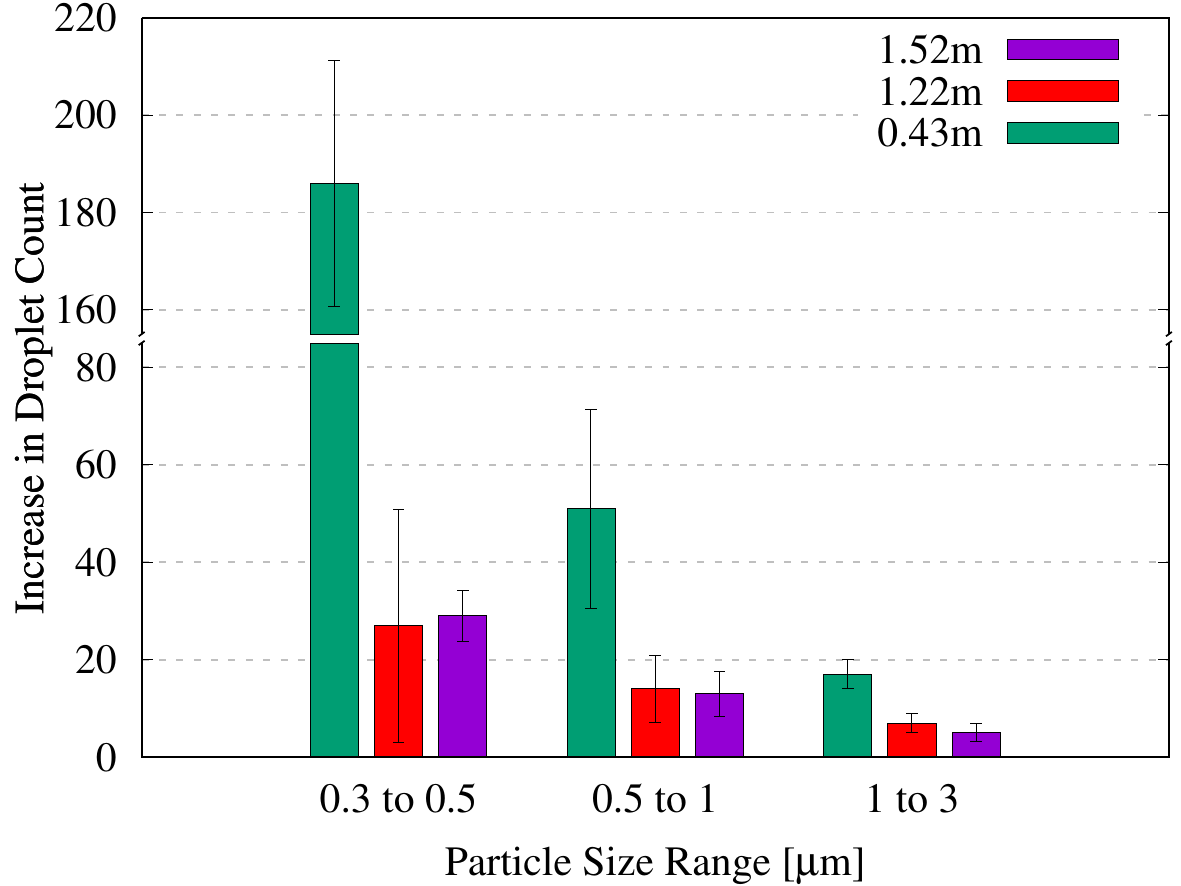}
	\caption{\label{fig:UnToiletDelta}Average increase in the number of droplets measured per second after flushing the toilet. \blue{The error bars indicate the standard deviation of the measured increase from multiple flushes.} Each bar cluster corresponds to particles in a given size range, and indicates how the droplet count varies with measurement height. The corresponding values are provided in Table~\ref{tab:ToiletTable}.}
\end{figure}

The number of droplets produced during the flushes were determined by numerically integrating the sections comprising the `surge' segments in the unfiltered time series. More specifically, within each 1-minute window associated with a particular flush, the start of the surge was identified using the time-delay values specified in Table~\ref{tab:startAndSpan}. Starting at this time the area under the particle-count curve was computed numerically up until the end of the surge, the corresponding time span for which is also specified in Table~\ref{tab:startAndSpan}. \blue{The average surge count was determined by dividing this area by the corresponding time span.} The area under the remaining parts of the curve, i.e., the segments lying outside the surge but within the 1-minute time window, was determined similarly to obtain the average ambient droplet count. This ambient count was subtracted from the surge count to yield the average number of flush-generated droplets measured per second by the particle counter. \blue{The resulting values from the 4 different full-minute flush measurements were averaged to obtain the increase in droplet count per second, and the standard deviation was computed to determine the uncertainty.} The resulting data for the flushing toilet is depicted graphically in Figure~\ref{fig:UnToiletDelta}, and the corresponding numerical values are provided in Table~\ref{tab:ToiletTable}.
\begin{table}[htp]
	\caption{\label{tab:ToiletTable}Numerical values for the average increase in droplet count per second from the toilet-flushing tests\blueTwo{, with the standard deviation provided in parentheses}. The data corresponds to the bar graphs shown in Figure~\ref{fig:UnToiletDelta}.}
\blue{
\begin{ruledtabular}
\begin{tabular}{lccc}
Height & (0.3 to 0.5)$\mu m$ 	&(0.5 to 1)$\mu m$  &(1 to 3)$\mu m$  \\ \hline
	$0.43m$  	& 186 ($\pm 25$)		& 51 ($\pm 20$)		& 17 ($\pm 3$)	\\
$1.22m$  		& 27 ($\pm 24$)		 	& 14 ($\pm 7$)		& 7 ($\pm 2$) 		\\
$1.52m$       	& 29 ($\pm 5$)		& 13 ($\pm 5$)		& 5 ($\pm 2$)	
\end{tabular}
\end{ruledtabular}
}
\end{table}
We note that droplets larger than 3$\mu m$ were excluded from this analysis since very few droplets in these size ranges were detected at the higher locations, which made it difficult to distinguish between the measured values and \blue{background} noise.

The bar graphs in Figure~\ref{fig:UnToiletDelta} indicate that a significant number of droplets smaller than $0.5 \mu m$ were generated by the flushing toilet. If these droplets contain infectious microorganisms from aerosolized biomatter, they can pose a significant transmission risk since they remain suspended for long periods of time. For instance, in a poorly ventilated location where gravitational settling is the only means of removing suspended particles, the Stokes settling time for a spherical water droplet of size $0.5 \mu m$ from a height of $1.52m$ ($5ft$) would be approximately 56 hours, or more than 2 days. Apart from the smallest aerosols, comparatively larger aerosols also pose a risk in poorly ventilated areas even though they experience stronger gravitational settling. They often undergo rapid evaporation in the ambient environment and the resulting decreases in size and mass, or the eventual formation of droplet nuclei, can allow microbes to remain suspended for several hours\blue{~\cite{Wells1934,Duguid1946,Basu2020}}. 
 
\blue{In Figure~\ref{fig:UnToiletDelta}, we observe a large variation for aerosols in the size range (0.3 to 0.5)$\mu m$. This may be attributed to the small droplets' high sensitivity to ambient flow fluctuations, as well as to the sensor's limited counting efficiency in this range. Notably, droplets smaller than $3\mu m$ are detectable in significant numbers even at a height of $1.52m$ ($5ft$). We observe a consistent decline in droplet count with increasing height; there is a significant drop in droplet count going from seat-level to $1.22m$, and a very small decrease with a further move up to $1.52m$ for droplets larger than $0.5\mu m$. The smallest aerosols exhibit some variation in the trend, which is likely due to the sensor limitations mentioned above. The observed decrease in droplet count with increasing measurement height} is expected, since the {droplet concentration is highest when the probe is placed closer to the flushing water, and it decreases at farther locations} due to dispersal of the droplets over a wider area. We remark that gravitational forces \blue{are not expected to} play a dominant role in the observed behavior, given the extremely small mass of the aerosols \blue{being considered here.} \blueTwo{Rather, it is aerodynamic drag that dominates.} {The Stokes settling speed for the largest aerosol being considered, i.e., a $3\mu m$ droplet, is approximately $0.00027m/s$. This amounts to a settling time of $1589s$ from a height of $0.43m$, and even longer for the smaller droplets. Thus, the effects of gravitational forces are not dominant at the time scales being considered ($\sim O(10s)$).} \blueTwo{Finally, the monotonic decrease in particle count with increasing particle size is similar to the trend observed by Johnson et al.~\cite{Johnson2013Aerosol} for various toilet designs and flushing mechanisms.}

\begin{figure}[h!]
\centering
\includegraphics[width=\linewidth]{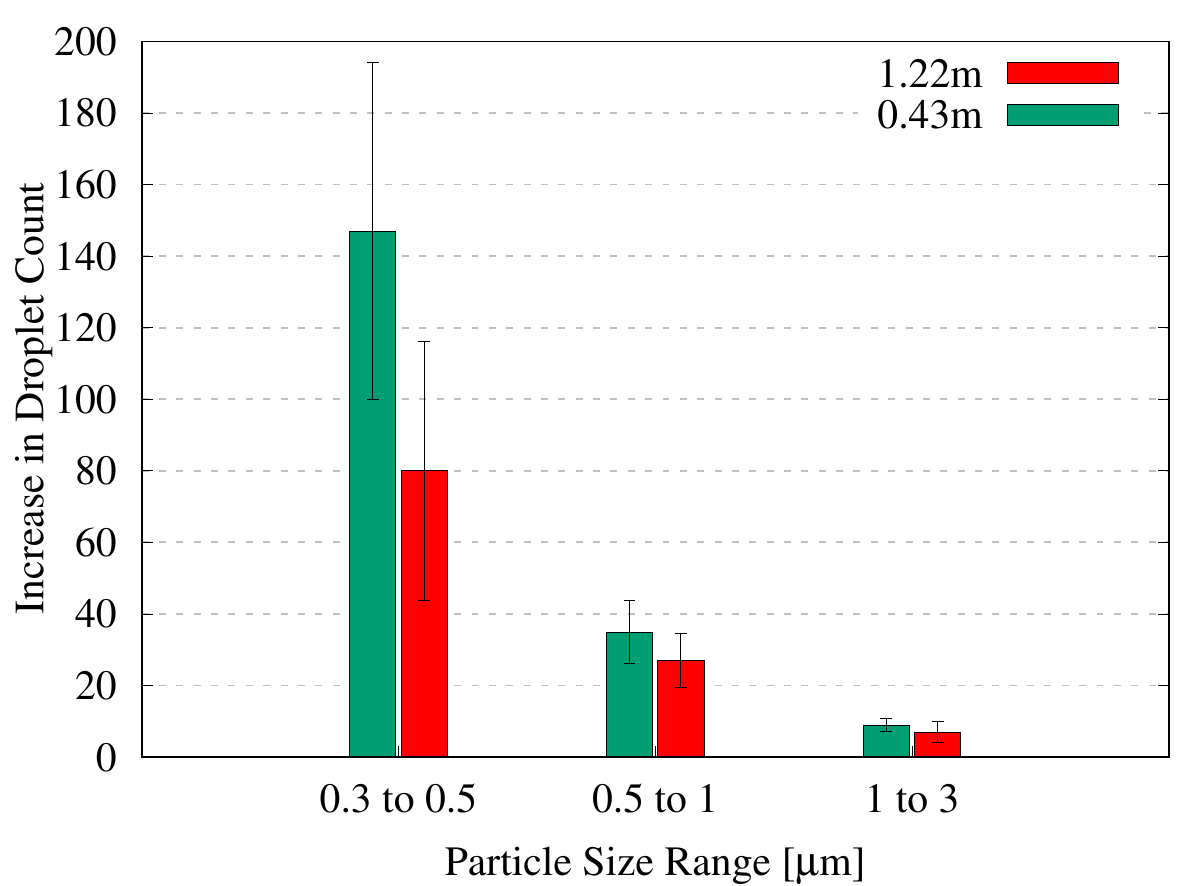}
\caption{\label{fig:CovToiletDelta}Average increase in number of droplets measured per second from flushing the covered toilet. \blue{The error bars indicate the standard deviation of the measured increase from multiple flushes.} Each bar cluster corresponds to particles in a given size range, and indicates how the droplet count varies with measurement height. The corresponding values are provided in Table~\ref{tab:CovToiletTable}.}
\end{figure}

\begin{figure}[h]
\centering
\includegraphics[width=\linewidth]{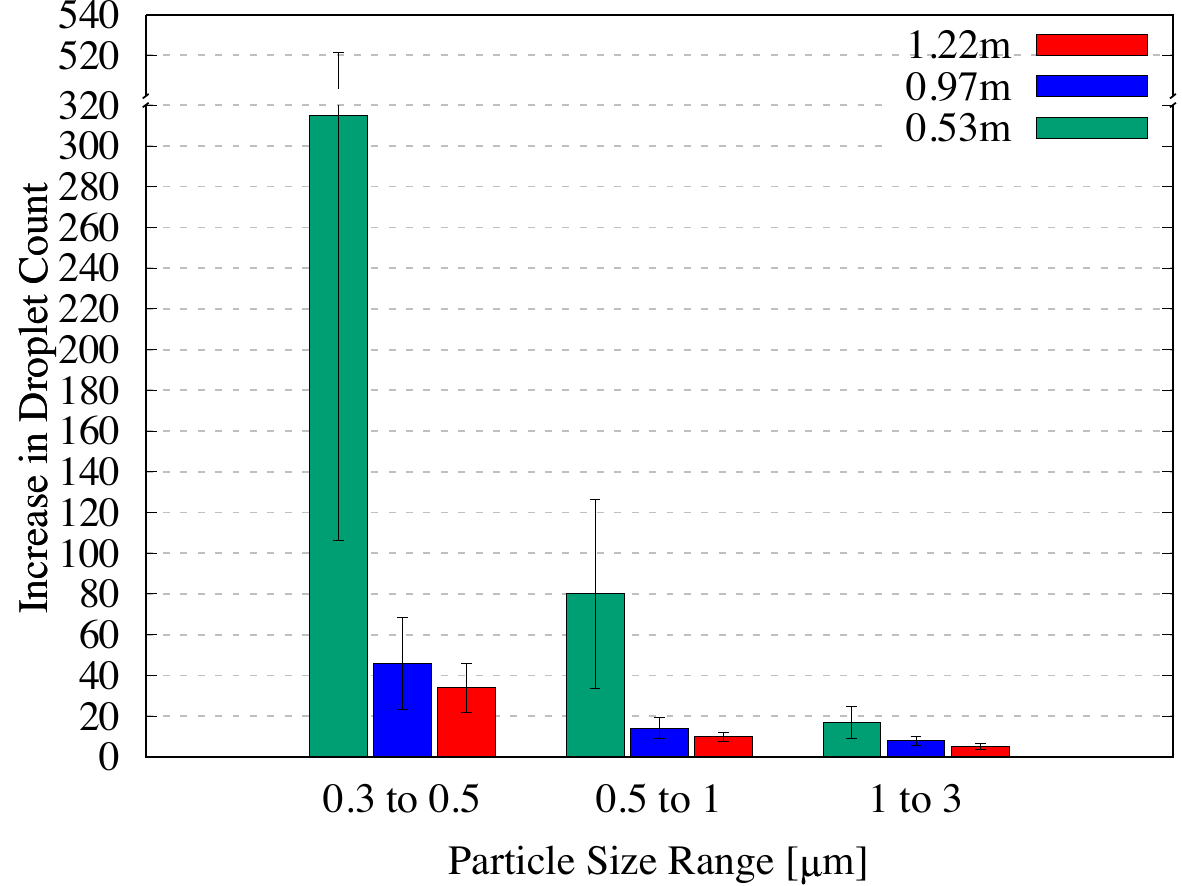}
\caption{\label{fig:UrinalDelta}Average increase in number of droplets measured per second from flushing the urinal. \blue{The error bars indicate the standard deviation of the measured increase from multiple flushes.} Each bar cluster corresponds to particles in a given size range, and indicates how the droplet count varies with measurement height. The corresponding values are provided in Table~\ref{tab:UrinalTable}.}
\end{figure}

\blue{The data collected after flushing the covered toilet and the urinal were also processed in a similar manner to determine the corresponding increases in droplet count per second. The results for the covered toilet} are presented in Figure~\ref{fig:CovToiletDelta} and Table~\ref{tab:CovToiletTable}, whereas those from \blue{flushing the urinal} are presented in Figure~\ref{fig:UrinalDelta} and Table~\ref{tab:UrinalTable}. Results from measurements at $1.52m$ height were not included in the analysis for the covered toilet, since \blue{it was difficult to discern droplet counts from background noise due to the extremely low measured values}. This indicates that the covering plate prevented the aerosols from rising upward \blue{and instead deflected them to lower levels, also resulting in shorter time delays compared to the uncovered toilet (Table~\ref{tab:startAndSpan})}. Over the long term however, these aerosols could rise up with updrafts created by the ventilation system or by \blue{the movement of people in the restroom}.

We observe a large number of aerosolized droplets smaller than 1$\mu m$ in Figure~\ref{fig:CovToiletDelta}, and an appreciable number of droplets in the (1 to 3)$\mu m$ range. \blue{This suggests that while the covering is able to suppress the dispersion of droplets to some extent, it does not eliminate them completely. Thus, although a toilet lid may appear to be a straightforward solution for reducing aerosol dispersal, other alternatives may need to be evaluated when designing public restrooms, such as modifying the fixture design, water pressure, vent placement, airflow rate, or even employing a liquid `curtain' incorporated into the fixture~\cite{Wu2020Curtain}.} 
\begin{table}[htp]
\caption{\label{tab:CovToiletTable}Numerical values for the average increase in droplet count per second from the covered toilet-flushing tests\blueTwo{, with the standard deviation provided in parentheses}. The data corresponds to the bar graphs shown in Figure~\ref{fig:CovToiletDelta}.}
\blue{
\begin{ruledtabular}
\begin{tabular}{lccc}
Height & (0.3 to 0.5)$\mu m$ 	&(0.5 to 1)$\mu m$  &(1 to 3)$\mu m$  \\ \hline
	$0.43m$  	& 147 ($\pm 47$)		& 35 ($\pm 9$)		& 9 ($\pm 2$)	\\
$1.22m$  		& 80 ($\pm 36$)		& 27 ($\pm 7$)		& 7 ($\pm 3$)		
\end{tabular}
\end{ruledtabular}
}
\end{table}

\blue{The bars in Figure~\ref{fig:CovToiletDelta} display a consistent decline in droplet count with increasing height, similar to the trend observed for the uncovered toilet. One unexpected observation is the occurrence of higher droplet counts for the covered toilet at $1.22m$, compared to analogous measurements for the uncovered toilet in Table~\ref{tab:ToiletTable}. We remark that this does not indicate that the covering led to an increase in droplet count, but rather that the aerosols were redirected in higher concentrations to the position where the counter was located, after being forced through gaps between the seat and the cover. Examining the data from the the urinal-flushing tests in Figure~\ref{fig:UrinalDelta}, we observe a similar decline in droplet count with increasing height as for the other two cases. A large number of droplets were detected in the (0.3 to 0.5)$\mu m$ size range (approximately 300 droplets per second on average) at the lowest measurement level, which can be attributed to the close proximity of the sensor to the flushing water. Moreover, a significant number of droplets reached heights of up to $1.22m$ ($4ft$) from the ground, similar to the toilet-flushing tests.}
\begin{table}[htp]
\caption{\label{tab:UrinalTable}Numerical values for the average increase in droplet count per second from the urinal-flushing tests\blueTwo{, with the standard deviation provided in parentheses}. The data corresponds to the bar graphs shown in Figure~\ref{fig:UrinalDelta}.}
\blue{
\begin{ruledtabular}
\begin{tabular}{lccc}
Height & (0.3 to 0.5)$\mu m$ 	&(0.5 to 1)$\mu m$  &(1 to 3)$\mu m$  \\ \hline
	$0.53m$  	& 315 ($\pm 209$)	& 80 ($\pm 47$) 		& 17 ($\pm 8$)		\\
$0.97m$  & 46 ($\pm 23$)		& 14 ($\pm 5$)			& 8 ($\pm 2$)		\\
$1.22m$ & 34 ($\pm 12$)		& 10 ($\pm 2$) 		& 5 ($\pm 2$)
\end{tabular}
\end{ruledtabular}
}
\end{table}

\blue{We remark that the total number of droplets generated in each flushing test described here can range in the tens of thousands. The numbers reported here indicate average droplet count per second, for cases where the time span for each surge varies from $15s$ to $20s$ (Table~\ref{tab:startAndSpan}). Thus, an average count of 50 droplets per second for one size range would amount to a total of 750 to 1000 droplets at one particular measurement location. Considering that similar measurements could be made all around the periphery of the fixtures, and that droplets are generated in several different size ranges, the overall total count would likely end up being significantly higher.} \blueTwo{Furthermore, droplet generation and accumulation depend on a variety of factors, such as the design of the toilet fixtures, the water pressure, the ventilation positioning, airflow, temperature, and RH, to name a few. The aim of the present work is not to present detailed characterizations of the influence of these factors on droplet dynamics, but instead to highlight the occurrence of aerosol generation and accumulation within public restrooms. These observations can help stimulate further studies to investigate steps to mitigate the issues involved.} {We further note that while the results presented here are restricted to specific measurement heights, there is a high likelihood of the aerosols getting dispersed throughout the room over time due to updrafts created by the ventilation system or by the movement of people.} 

\begin{figure}[h]
\centering
\includegraphics[width=\linewidth]{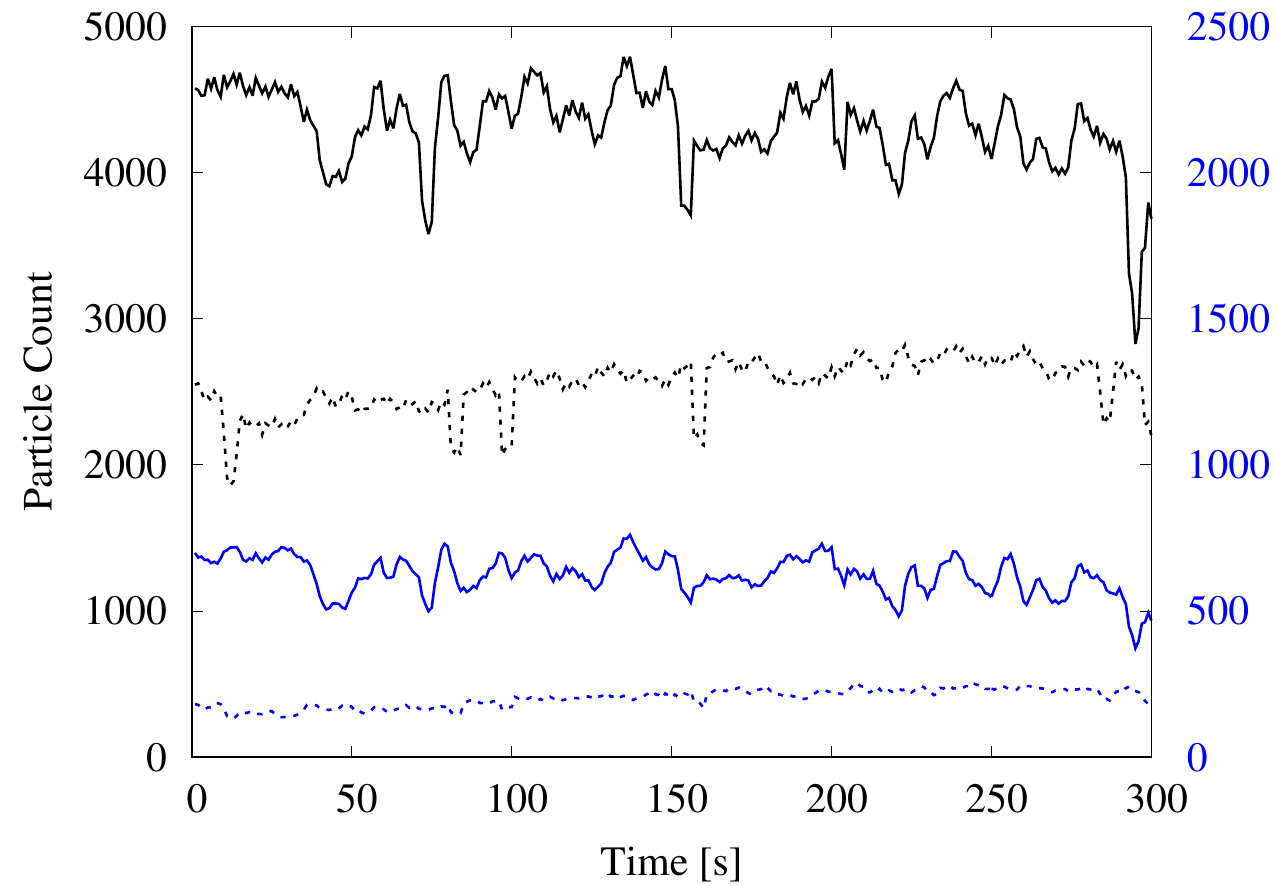}
\caption{\label{fig:BackgroundChange} Particle-count from ambient measurements within the restroom. The plot indicates the time-variation of particles in two different size ranges, (0.3 to 0.5)$\mu m$ - black, and (0.5 to 1)$\mu m$ - blue. The black curves correspond to the left vertical axis, whereas the blue curves correspond to the right vertical axis. The dashed lines indicate initial background readings before conducting any flushing tests, whereas the solid lines indicate measurements taken at the conclusion of all tests, approximately 3 hours and 100 flushes later.}
\end{figure}

\begin{figure}[h]
\centering
\includegraphics[width=\linewidth]{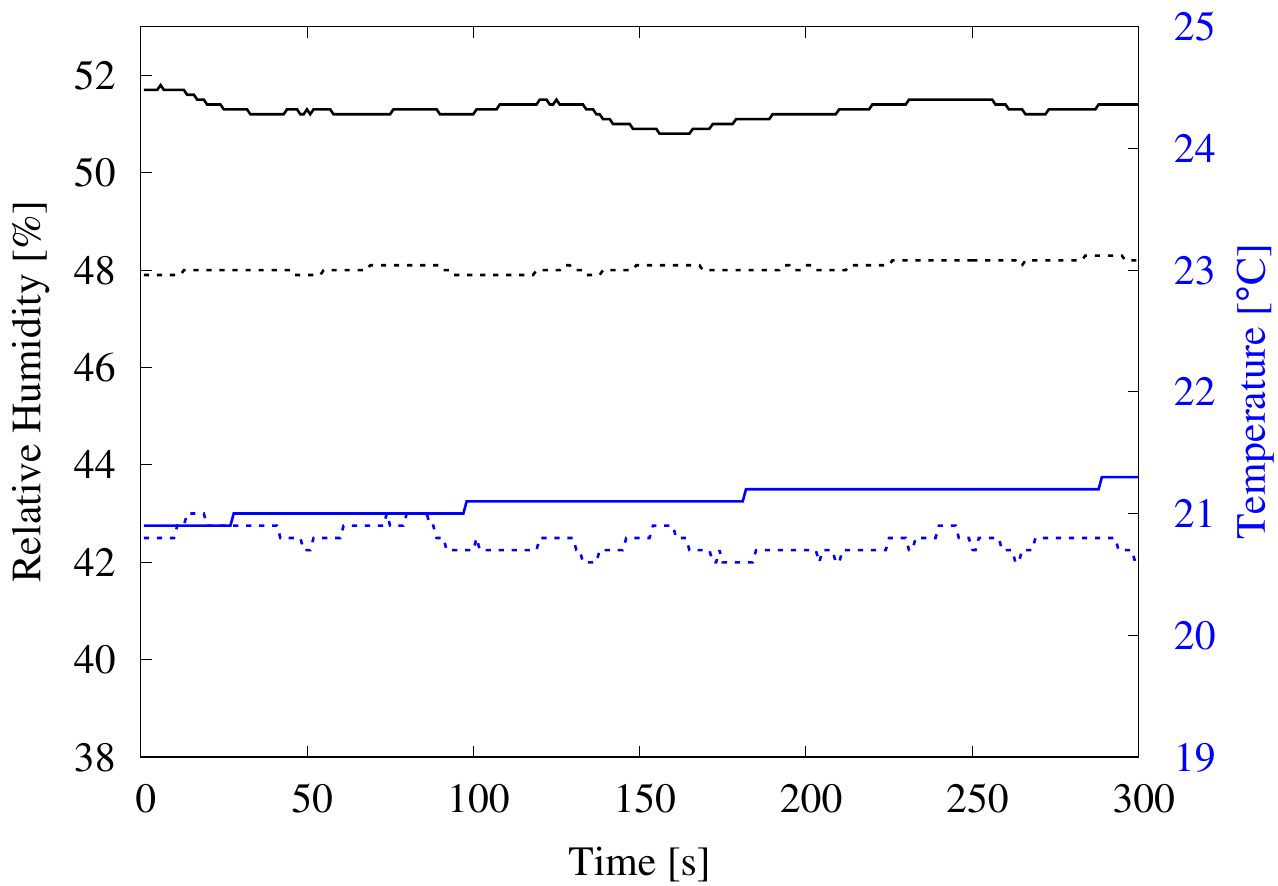}
	\caption{\label{fig:RH_temp} \blueTwo{Relative humidity (black) and temperature measurements (blue) inside and outside the restroom. Solid lines indicate measurements taken inside the restroom, whereas dashed lines correspond to measurements outside the restroom, a few meters away from the closed restroom door.}}
\end{figure}

In addition to the flush-generated aerosol measurements, ambient aerosol levels were measured prior to starting the experiments and again after completing all of the tests. After approximately 3 hours of tests involving over 100 flushes, there was a substantial increase in the measured aerosol levels in the ambient environment. The corresponding data is presented in Figure~\ref{fig:BackgroundChange} and Table~\ref{tab:BackgroundValues}.  
\begin{table}[htp]
	\caption{\label{tab:BackgroundValues} Average values for the background measurements shown in Figure \ref{fig:BackgroundChange}. Additionally, average measurements for the (1  to 3)$\mu m$ size group are also provided below. The `Before' column indicates the average ambient levels measured within a 5-minute time window before conducting any flushing experiments, and the `After' column indicates similar measurements taken after concluding all the experiments.}
\begin{ruledtabular}
\begin{tabular}{crrr}
Particle Size Group & Before & After  & Percent Change \\ \hline
0.3 to 0.5 $\mu$m   & 2537 & 4301 	& 69.5\%          \\
0.5 to 1 $\mu$m     	& 201 	& 621 		& 209\%         \\
1 to 3 $\mu$m       	& 8   	& 12  		& 50\%
\end{tabular}
\end{ruledtabular}
\end{table}
There was a $69.5\%$ increase in measured levels for particles of size (0.3 to 0.5)$\mu m$, a $209\%$ increase for the (0.5 to 1)$\mu m$ particles, and a $50\%$ increase for the (1 to 3)$\mu m$ particles. Particles larger than 3$\mu m$ were excluded from the analysis due to the the impact of background noise on the extremely low measured values. The results point to significant accumulation of flush-generated aerosolized droplets within the restroom over time, which indicates that the ventilation system was not effective in removing them from the enclosed space, although there was no perceptible lack of airflow within the restroom\blue{; the room was equipped with two vents rated at volume flow rates of $7.5 m^3/min$ (265 CFM) and $5.66 m^3/min$ (200 CFM)}. Furthermore, a comparison with ambient levels outside the restroom (a few meters away from the closed restroom door, but within the same building) indicated that the levels of droplets smaller than $1\mu m$ were more than 10 times higher within the restroom compared to ambient levels outside the restroom. This was unexpected since the restroom had been closed off for more than 24 hours after deep cleaning, with the ventilation system operating normally. \blue{While it is difficult to ascertain the exact source of the droplets that contributed to high background levels within the restroom, it is likely that they were generated during the cleaning operation. There were no other readily apparent sources, since both locations, i.e., inside and outside the restroom, employed the same centralized air-conditioning system, and the RH and temperature were maintained at comparable levels \blueTwo{(Figure~\ref{fig:RH_temp})}.} These observations \blue{further} highlight the importance of employing adequate ventilation in enclosed spaces \blue{to extract suspended droplets effectively, in order to reduce} the chances of infection transmission via aerosolized droplets.
 
The results presented here indicate that although the likelihood of infection for respiratory illnesses via bioaerosols may be low compared to the risk posed by respiratory droplets (since virions are detected in larger quantities in respiratory samples), it presents a viable transmission route especially in public restrooms which often experience heavy foot-traffic within a relatively confined area. As demonstrated here, multiple flush-use over time can lead to an accumulation of potentially infectious aerosols, which poses a measurable risk considering the large number of individuals who may visit a public restroom and subsequently disperse into the broader community. Moreover, apart from flush-generated bioaerosols, the accumulation of respiratory aerosols also poses a concern in public restrooms if adequate ventilation is not available. \blue{Overall, the results presented here highlight the crucial need for ensuring effective aerosol removal capability in high density and frequently visited public spaces.}

\section{Conclusion}
\label{sec:conclusion}
The aerosolization of biomatter from flushing toilets is known to play a potential role in spreading a wide variety of gastrointestinal and respiratory illnesses. To better understand the risk of infection transmission that such droplets may pose in confined spaces, this paper investigates droplet-generation by flushing toilets and urinals in a public restroom operating under normal ventilation condition. The measurements were conducted inside a medium-sized public restroom, with a particle counter placed at various heights to determine the size and number of droplets generated upon flushing. The results indicate that  both toilets and urinals generate large quantities of droplets smaller than 3$\mu m$ in size, which can pose a significant transmission risk if they contain infectious microorganisms from aerosolized biomatter. The droplets were detected at heights of up to $1.52m$ ($5ft$) for 20 seconds or longer after initiating the flush. Owing to their small size, these droplets can remain suspended for long periods of time, as is demonstrated in the present study via ambient measurements taken before and after conducting the experiments. When a large flat plate was used to cover the toilet opening, it led to a decrease in droplet dispersion but not a complete absence of the measured aerosols. This indicates that installing toilet seat lids in public restrooms may help reduce droplet dispersal to some extent, but it may not sufficiently address the risk posed by the smallest aerosolized droplets. Ambient aerosol levels measured before and after conducting the experiments indicated a substantial increase in particle count, pointing to significant accumulation of flush-generated aerosols within the restroom over time. This indicates that the ventilation system was not effective in removing the aerosols, although there was no perceptible lack of airflow within the restroom. Importantly, this suggests that multiple flush-use over time can lead to the accumulation of high levels of potentially infectious aerosols within public restrooms, which poses an elevated risk of airborne disease transmission. In addition to flush-generated bioaerosols, the accumulation of respiratory aerosols also poses a concern in public restrooms in the absence of adequate ventilation. Overall, the results presented here indicate that ensuring adequate ventilation in public restrooms is essential, since these relatively confined areas often experience heavy foot traffic and could pose a risk for widespread community transmission of various gastrointestinal and respiratory illnesses.

\section*{Data Availability}
The data that support the findings of this study are available from the corresponding author upon reasonable request.

\bibliography{toiletPaper}

\end{document}